# Beyond journals and peer review: towards a more flexible ecosystem for scholarly communication

*Michael Wood*
*University of Portsmouth, UK*
*michael.wood@port.ac.uk or michaelwoodslg@gmail.com*

## Abstract

This article challenges the assumption that journals and peer review are necessary for developing, evaluating and disseminating scientific and other academic knowledge. It suggests a more flexible ecosystem and examines some of the possibilities this might facilitate. The market for academic outputs should be opened up by encouraging the separation of the dissemination service from the evaluation service. Publishing research in subject-specific journals encourages compartmentalising research into rigid categories. The dissemination of knowledge would be better served by an open access, web-based repository encompassing all disciplines, or by a number of such repositories. There would then be a role for organisations to assess the research items to help users find relevant, high-quality work. There could be a variety of such organisations which could enable reviews from peers to be supplemented with evaluation by non-peers from a variety of different perspectives: user reviews, statistical reviews, reviews from the perspective of different disciplines, and so on. This should reduce the inevitably conservative influence of relying on two or three "peers", and make the evaluation system more critical, multi-dimensional and responsive to the requirements of different audience groups, changing circumstances, and new ideas. Non-peer review might make it easier to challenge dominant paradigms, and expanding the potential audience beyond a narrow group of peers might encourage the criterion of simplicity to be taken more seriously - which is essential if human knowledge is to continue to progress.

## Introduction

Papers published in peer reviewed journals are a key part of the process by which academic knowledge is developed. Articles are checked through by "peers" to try to "ensure that the valid article is accepted, the messy article cleaned up, and the invalid article rejected" (Weller, 2001: p. xii). The goal of this process is to ensure that the papers published are of as high quality as possible, so that fellow researchers and other readers can have confidence in what they read.

According to Larsen and von Ins (2010) the number of peer reviewed journals "most likely is about 24,000". Authors of research articles wishing to publish in one of these journals, must first select their journal (sometimes not easy, given the choice), make sure their article conforms to the requirements of the chosen journal, send a copy to the editor who will decide if it is worth serious consideration, in which case he or she will send it out to two or three peer reviewers, and then make a decision based on the recommendations of these reviewers. This decision might be to publish the article as it is, to reject it, or to ask the authors to improve the article in line with the reviewers' recommendations and then reconsider it. The reviewers are "peers" in the sense that they are typically working in the same area, and may have submitted work to the journal in question. They are usually unpaid, anonymous volunteers who cannot be expected to respond immediately, so the whole process may take some time - in the worst cases, several years. The details of this process vary from journal to journal, and conventions in different disciplines differ, but the above outline is typical. However, this is a relatively recent development: "Peer review in its modern present form is only about 40 years old and is not standardized. ... a systematic peer review for Nature was only introduced in 1966 ... Proceedings of the National Academy of Sciences introduced peer review only a few years ago" (Larsen and von Ins, 2010).

There are differences between disciplines, but in general far more credence is placed on research published in peer reviewed journals than on other sources: researchers typically only take work published in peer reviewed journals seriously, students are routinely told that the only references that count are to peer reviewed journals, and when research results are given in news bulletins the name of the journal is often given as evidence that they are worth taking seriously. This is not to deny, of course, that there are other useful sources for researchers – blogs from respected sources, books, articles in newspapers and magazine, and so on – but my concern in this article is the peer reviewed journal.

Although the peer review journal system is still the gold standard in most disciplines, it has attracted considerable criticism. The most commonly made points are:

1. Most journals are expensive for readers to access which, in practice, means that the readership is often restricted to those with university subscriptions to the journals. This has led to calls for open access journals, which are being increasingly heeded. The system does seem to be progressing in the direction of open access so that the products of research should be accessible to all.
2. The fact that there are typically only two or three volunteer reviewers chosen by the editor, may lead to the suspicion that "bad" papers may sometimes be published and "good" papers rejected. Empirical studies sometimes find little correspondence between the verdicts of different reviewers (see references below), so which papers get to be published may be little more than a lottery. This means that peer reviewed journals may not be an efficient quality monitoring system.

3 The demand for quality assessment of research has led to quality ranking of journals so that the journal a paper appears in can be used to assess the quality of the research reported in a paper. This has led to the feeling that the top ranking journals are unlikely to accept risky or innovative papers for fear that this may damage their status. This may have a detrimental effect on the growth of knowledge, and encourage researchers to write papers according a formula approved by the top journals.

4 It may take a long time to get a paper published, particularly if it fails to be accepted by the first choice journal and is submitted to a sequence of journals, each of which may take months, or even years, to make a decision (parallel submissions are not allowed). In a study of one author's submissions over a given time period, the average time between submission and appearance in print for the 39 articles which had been published was 677 days (nearly two years); he had a further 14 articles he was still trying to find an outlet for - on the date he did the analysis these articles had been waiting an average of 3.5 years since their initial submission and had been submitted to between 1 and 5 journals (Nosek and Bar-Anan, 2012). My experience[1] suggests that this is not untypical. This has obvious implications in terms of slowing down the growth of knowledge, as well as being frustrating for researchers and their audience.

Appendix 1 below contains sketches of a few stories which illustrate some of the problems of the current system.

These problems are serious. The Covid 19 crisis has provided an incentive to speed up the peer review process which has proved that this is often possible, but it has also highlighted the warning "not yet peer reviewed" about papers posted on repositories like medrxiv.org. This suggests, firstly, that papers on such repositories are not reviewed or checked by anyone, and secondly, that once a paper has been peer reviewed its conclusions are beyond doubt: both of these assumptions are far too simplistic.

In an earlier era, Einstein's papers on special and general relativity were accepted by the journals he submitted them to (although without peer review according to Adler, 2012)[2], but many such revolutionary ideas may have been rejected. If this had happened to Einstein's papers, physics may have progressed more slowly or along a different route. Or perhaps, someone else came up with ideas similar to Einstein's years before him? There is no way of knowing: counterfactual history is a tricky discipline. Gregor Mendel's now famous work on genetics was published in 1866 but was rarely cited and Darwin appears not to have known about it (Fairbanks, 2020) although it was undoubtedly very relevant to his ideas on evolution. The publication system had failed to adequately distribute a theory which subsequently turned out to be very useful. Many other useful ideas have almost certainly never seen the light of day. Although we obviously cannot point to these ideas, it is possible to model the possible impact of the inadequacies of a publication system: see Appendix 2 below.

[1] My record for the time delay between submission and publication is 4 years and 4 months.

[2] In 1905, when his first revolutionary papers were published, he was unknown and did not have an academic job, so it would have been understandable if the journals had not given serious consideration to his work.

Not surprisingly, there is an extensive literature on these problems and how they might be remedied: e.g. Tennant and Ross-Hellauer (2020), Schmidt and Görögh (2017), Kovanis et al (2017), Adler (2012), Nosek and Bar-Anan (2012), Priem and Hemminger (2012), Bornmann (2011), Weller (2001). Experimental journals and peer review and publishing platforms incorporating new ideas include Cureus (http://www.cureus.com/), InterJournal (http://www.interjournal.org/), PubPeer (https://pubpeer.com), Principia (http://www.principia.network), "linked research" (https://linkedresearch.org/), Scholastica (https://scholasticahq.com/), Qeios (https://www.qeios.com/) and the "alternative review tools and services" listed in Table 1 of Schmidt and Görögh (2017). “Plan S” (see Liverpool, 2023) is a proposal by a “group of influential funding agencies, called cOAlition S” to overhaul the whole system so that “all versions of an article and its associated peer-review reports [would be] published openly from the outset, without authors paying any fees, and for authors, rather than publishers, to decide when and where to first publish their work … publishers are no longer gatekeepers that reject submitted work or determine first publication dates”.

This literature provides a number of suggestions for making peer review quicker and more effective - e.g. reviews from the first choice journal being passed on to another journal if the paper in question is not accepted by the first choice journal, post publication reviews, open (published) reviews, and so on. However, most of these suggestions start from the premise that journals and peer review are necessary and sufficient.

The perfect system would deliver to every relevant person just those artefacts that are useful and of a sufficiently high quality, and do so immediately with no publication delay. Merely stating this makes it clear that this ideal is both impractical, and not clearly defined (how are usefulness and quality defined from the perspective of each person?). However, imperfections are easier to spot (see Appendix 1): it seems plausible that even a small improvement could make a large difference. Perhaps, for example, medical advances which have not yet happened might already be common practice with a more efficient system for distributing knowledge?

If there were a universally accepted checklist against which a new research article could be evaluated in an objective way, so that all reasonable people would reach the same conclusion, then the standard peer review system might be efficient and reliable. In practice, in most disciplines, this is far from the case: the process of reviewing a research paper requires subjective judgments and differences of opinion are likely. It is impossible to predict which research efforts the future will judge favourably: we can only guess. Given this, it seems, to me, almost inevitable that the present system will be arbitrary and, given the high stakes involved for the careers of academics, on occasions, possibly corrupt.

The current system depends on two key components: *journals* – most of which are *subject-specific,* and *peer-review*. In the next two sections I discuss these two concepts and argue that they are both, to some extent, problematic. I then discuss two principles which could lead to a very different ecosystem with many advantages over the current one. The later sections of the article explore some of these advantages, and also some of the barriers to change.

It is important to mention three preliminary issues. The first is the unit of research output. Traditionally this is the "paper" but there are strong arguments that this should be broadened and the name changed. In the UK the term "research output" is sometimes used in documentation relating to research assessment. In this paper I will use the phrase "academic artefact" to refer to artefacts of potential value produced by academics, such as research papers (which, of course, are no longer usually printed on paper), or, for example, videos or data sets. Some further issues concerning academic artefacts are outlined in Appendix 3.

Second, academia encompasses many different disciplines and often different "tribes" within a discipline. This means that there is a great variety of types of research including, for example, randomized control trials in medicine, case studies in management, conceptual analysis in philosophy, and symbolic arguments in mathematics, to mention just a few. Words like "research", "scholarly", "scientific" and "rigorous" may seem natural in some contexts but not others and may have different implications in different contexts. The use of the word "scholarly" in the title is intended to signal that my intention in this article is to provide a general analysis, not tied to a specific discipline.

Third, my aim in this article is to analyse some problems with the current system and look at *possible* solutions. These are just possibilities or suggestions: I obviously cannot be sure they will work, but all change must start by envisaging alternatives. I am not advocating, for example, abolishing peer review, but in appropriate circumstances supplementing it by other types of review. Then, in due course, innovations which are found to be useful will, hopefully, spread.

## Subject-specific journals

A few years ago journals were printed on paper, often by learned societies, and distributed to members of the society and libraries. Researchers in a particular discipline would browse through the appropriate journals to find material relevant to their research. They would probably have the most relevant ones in their office, and the university library could supply the rest. Gradually the number of journals increased, indexing services arose to help researchers cope with a much larger search domain, and online availability means that most articles are now instantly available (but only to readers with a paid subscription in the case of journals which do not allow open access). Google Scholar is now probably the most used indexing service, and the typical researcher's tactics have changed from browsing key journals to searching on Google Scholar or other databases. The particular journal a paper is published in matters little now to researchers or their audience, but the prestige of the journal may be important to legitimise the research and persuade people to take it seriously.

However, subject specific journals persist. If I want to publish an academic article which wins the respect of the academic community I have to do so in a journal. This seemingly innocuous fact has many unfortunate consequences. I am only allowed to submit to one journal[3] at a time so if the first rejects it I may resubmit to a second, and so on - this can result in delays of years. (In many commercial domains this would be outlawed as a serious inefficiency in the market.) My paper may be interdisciplinary and it may not be obvious which journal is appropriate. If my paper presents, say, statistical evidence on a medical matter, I may want my work checked by both the statisticians and the medics. There are a few journals which cater for a broader range of subjects - e.g. *Sage Open* spans "the full spectrum of the social and behavioral sciences and the humanities" - but these are the exception not the rule.

## Peer review

Most of the published research on peer review is about its efficiency – does the current system provide valid and reliable assessments of research papers? The evidence here varies from discipline to discipline but is often negative. Peer review may be unreliable in the sense that different reviewers may, and often do, give different opinions, so taking just two or three reviewers makes the review process into a lottery - e.g. Kravitz et al (2010), Wood et al (2004), Peters and Ceci (1982). This is not surprising - two or three anonymous, unpaid, and unaccountable reviewers, who are probably just experts in some of the areas relevant to a thorough critique of the paper, would not be expected, on a priori grounds, to produce definitive reviews. Even when reviewers agree there is, of course, no easy way of deciding whether they are right, or even of defining what "right" means.

This does not mean, of course, that disagreement between reviewers should never happen. In the absence of unambiguous criteria for evaluating academic artefacts, disagreements are inevitable. Indeed, the interaction between different viewpoints is likely to be essential for the healthy growth of knowledge, and a good dissemination system should facilitate such interactions.

Besides recommending whether to accept or reject a paper, peer review is also useful as a source of comments and suggestions for improving the paper. I have recently received two reviews of a paper submitted to a peer reviewed journal: both reviews were positive in the sense that they thought a revised version of the paper should be published, but with one exception, the critical comments they made were completely different. This may well be typical and suggests that it may be helpful to have many reviews, certainly more than two, because the extra reviewers may raise additional issues.

The phrase "peer review" usually refers to peer review before publication with the aim of helping authors improve their paper and editors decide whether to publish it. The only public output from the peer review process is acceptance of the (possibly revised) paper for publication. More detailed comments about the paper, perhaps some of the reviewers' reservations, or their comments on strong points, or reasons for rejection of papers that are not published and may never see the light of day. This seems a lost opportunity. Open peer review means that more details of the review are available to readers, and post-publication peer review is possible but obviously cannot have a bearing on the decision to publish. However, these are not common practice in most fields. Shashok and Matarese

[3] For obvious reasons it is difficult for one journal to relax this restriction unilaterally, If it were possible for authors to submit their paper to a range of journals, and then accept the best publication offer, this would obviously transform the market for academic papers.

(2018) discuss both the value of post publication peer review, and the difficulties of publishing such reviews. Tennant and Ross-Hellauer (2020) give a detailed analysis of these, and other issues associated with peer review, and advocate "the development of a new research discipline based on the study of peer review."

However, there is another, more fundamental, problem with peer review: it is restricted to peers in the same discipline or sub-discipline. It is no longer considered acceptable that doctors, teachers, and so on, only review themselves: review by customers, and possibly other stakeholders, is now the norm. And, of course, producers of cars, haircuts, computers, and meals have always been subject to market forces: the ultimate test is whether people are prepared to spend their money. The academics' argument in defence of their exception from the general rule of external review is that only peers are in a position to review their work, and there is some justice in this. However, a system without any mechanism to provide feedback from potential customers or other audiences, including researchers in other areas, may fail to generate the most relevant and interesting research. Small cliques of academics may review each other's work positively despite its lack of interest or relevance to anyone else. Sometimes this may work, and the argument that short term responsiveness to customer feedback may inhibit progress may be valid, but as a general principle it seems suspect. There is a strong case for at least a limited amount of *non-peer review.* This category might also include non-standard, perhaps paradigm changing, views, within a discipline.

## What's the point of peer reviewed journals?

From the point of view of the consumers of, or audience for, academic artefacts, it is helpful to distinguish three purposes for peer reviewed journals:

1. They disseminate (some) academic artefacts to their audience.
2. They (supposedly) provide some sort of quality assurance that the artefacts are right, or true, or at least the best available at present.
3. The process of finding work on a particular topic is (supposedly) simplified by eliminating artefacts that are not up to scratch and so should be ignored.

Obviously, they are profitable for publishers, and they may provide a means of publicising research or exerting influence for academics and other stakeholders, but their primary purpose must be to communicate to their audience.

It is obvious that there are some difficulties in achieving these objectives with current peer reviewed journals. Some suspect work may be published in peer reviewed journals, and some useful work will not be published and so will be missed by the potential audience. Given how easy it is to post papers on repositories such as arxiv.org and SSRN, there now seems little point in peer reviewed journals apart from their quality assurance function, which could be done better by organizations with a more flexible remit as explained below.

## Two suggested principles for an academic ecosystem

Recent developments, and the literature on peer reviewed journals, suggest that the dissemination function of journals, and the evaluation function could be separated, and there may be substantial advantages in doing so. Priem and Hemminger (2012) recommend the "decoupled journal" in which "the functions [archiving, registration, dissemination, and certification] are unbundled and performed as services, able to compete for patronage and evolve in response to the market." The argument for this separation or decoupling is very similar to the arguments against excessive vertical integration in economic markets: it makes it easier for users to select the best option for each component in the process. Here, I will focus just on the dissemination and the evaluation (or review or assessment or certification) functions. Under this system authors or producers of academic artefacts would (1) deposit their paper or other artefact on a repository of some kind. It might then (2) be reviewed or evaluated by a review or evaluation service.

In the sections below I discuss the two principles underlying this system. With the expansion of open access journals and repositories, the first principle is already partially implemented, and the likelihood is that the system will evolve in the direction of all academic artefacts being feely available. The second principle would involve a change from existing practice, but as it would not involve any conflict with existing practices, it could be implemented gradually as suggested below. I will then go on to explore some of the possibilities these suggestions might facilitate.

## Principle 1. Academic artefacts should be freely available, with minimal categorisation and evaluation, and a transparent search mechanism

The ideal way of achieving this is to have a single global repository covering all disciplines (Wood, 2010). Let's call such a repository AcademicArtefacts.org (this website does not, of course, exist). At present, the majority of artefacts would be digital copies of documents (papers), but there might also be, for example, videos and datasets. In the case of artefacts which could not be deposited in the repository, there would be a link to where the artefact is located.

At present there are repositories for papers which are freely available but these are mostly either restricted to particular ranges of subjects (e.g. arxiv.org) or to a single institution. There are exceptions: for example Preprints (https://www.preprints.org/) which describes itself as "the multidisciplinary preprint platform". The name of this platform is based on the assumption that papers posted on them will go on to be submitted to peer reviewed journals, but this is not true of all papers posted, some of which might be rejected by peer reviewed journals and some of which might not be submitted. Qeios (https://www.qeios.com/) is another multidisciplinary platform which caters for both preprints and "open access papers". Qeios also has facilities for organising and encouraging peer reviews – see the next section for more on this.

There is an argument that subject specific repositories would be useful, because then people interested in, say, social enterprises, could go to the social enterprise repository. However, the difficulty is that there might be relevant material in general business repositories, or in repositories focusing on charities, and so on. Separate repositories for separate areas will lead to the same problem as that created by the 24,000 journals.

There is also an argument that if everything is on the web, search engines (such as Google or Google Scholar) can be used to search for material relevant to a given topic[4] so where the artefacts are stored does not really matter. There is some truth in this, but it is important to remember that web search engines are commercial products that use secret algorithms to order the results of a search. With a search which yields many hits (perhaps thousands or even millions) which ones appear at the top of the list will be determined by these secret algorithms which may be optimised for commercial purposes. This would not be true of a repository designed with its own search facility which is transparent to users (such as https://arxiv.org/).

In the case of Google Scholar the ordering of results is heavily, but not entirely, dependent on the number of citations to an article. In the case of Google the ranking is based on a less academic measure. As an example I tried searching on two keywords (quantum and Bayesian) on both sites and looked at the first five hits. There was no overlap in the documents found by the two searches; the citation counts for the first five hits on Google Scholar were: 269, 128, 69, 32, 49; and on Google were: no citation count given, 32, no citation count given, 4, 2. (The sites without citation counts were Wikipedia and Quanta magazine.)

Does it matter that these search engines use different, and secret, algorithms. I think it does because these algorithms will obviously influence what researchers read and base their research on. Nobody could read everything (Google Scholar claimed about 89,000 hits and Google about 4.8 million), so which results come near the top of the list is crucial. Do we want Google to set the research agenda even if it is more by accident than design?

But ... you may think that as a major criterion used by Google is citation counts for Google Scholar and another measure of popularity for Google, isn't this just a way of homing in on the "good" research? Again, there is doubtless some truth in this, but there are two reasons for caution. First, works with many citations are more likely to gather more citations just because they are at the top of the ranking. Even if all works were of equal value, those that by chance received more citations would rise in the rankings and receive even more. There will be a tendency for everyone to read the same works. Dissenting, alternative perspectives are unlikely to get a hearing. The second reason for caution is that citation counts are such an important measure for academics that they may be subject to manipulation.

This problem would be avoided, or much reduced, by AcademicArtefacts.org and bibliographic databases with a transparent search mechanism. It is obviously necessary to have some criterion for prioritising the artefacts (nobody could look at all 89,000 hits mentioned above), and citation counts might be one criterion for ordering results, but it would be possible to use other criteria, or even to choose works at random[5] - a possibility which is almost impossible with Google products. The important

[4] Gardner and Inger (2018) present the results of a survey which shows that the commonest approach to finding articles on a particular topic are specialist bibliographic databases, closely followed by general and academic search engines (p.9) - mainly Google and Google Scholar and Baidu in China (p.22). Among non academics it seems likely (to me) that the vast majority of searches would use one of the Google search engines (or Baidu in China).

[5] The potential advantage of choosing works at random is that it eliminates the influence of criteria which may downgrade helpful ideas. Imagine, for example, that one of Google Scholar's criteria is the number of references

thing is that the researcher (or other user) would have control of how the search results are ordered. And, of course, Google Scholar would doubtless include AcademicArtefacts.org so researchers would have the best of both worlds.

To take a specific example, Reimers, Knapp and Reimers (2012) conducted a

> systematic literature search ... in the electronic bibliographical database PubMed (http://www.ncbi.nlm.nih.gov/pubmed/). We only searched for English-language peer-reviewed journal articles using the search terms "(life expectancy OR longevity) AND (physical activity OR exercise OR sport)" ... A total of 1,932 articles were found.

Such a search would be much fuzzier using Google Scholar. Even this PubMed search inevitably excludes works not indexed by PubMed of which there are two important categories. First, works in areas not deemed medical by PubMed - longevity is an issue which is likely to be relevant to many diverse fields. Second, works which did not appear in peer reviewed journals because they were rejected or not submitted - an important subcategory of which is works which give negative results suggesting, for example, that an intervention does not work: these may be less likely than positive results to be deemed interesting and worthy of publication (e.g. see Nosek & Bar-Anan, 2012, p. 225). These categories of omissions mean that the literature survey is likely to be more biased than may initially be apparent. AcademicArtefacts.org, by encompassing all disciplines and not requiring peer review, should reduce the impact of these sources of bias.

There are also obvious advantages to any open web-based system. Documents on the web are far easier to obtain than old-style hard copies, and the page limit constraints of traditional paper journals should be a thing of the past - articles should no longer be turned down because there is no space in the journal and longer, more detailed articles would be feasible (Nosek and Bar-Anan, 2012). It is also far easier to publish updates and corrections where they are likely to be noticed, thus avoiding the problem that a printed journal article may continue to be accepted despite being discredited or outdated - see Appendix 1 for an example. On the other hand, there may be doubts about the longevity of some digital formats.

Any repository would need to need to impose some restrictions on material submitted to ensure that only "academic" artefacts are posted. However, my argument is that assessment or review beyond this minimal level, should only be carried out *after* an academic artefact has been posted in a public forum as explained in the next section.

## Principle 2. Organisations for assessing academic artefacts are necessary, and there are advantages in separating these from organisations for dissemination

cited in a paper. Papers citing no other works would come right down the list and probably be ignored, but it is possble that one such paper, coming at the problem from different perspective from all the other papers, may hold the key to progress. With a random choice such papers would have the same chance of appearing at the top of the list as all the other papers. In practice, a random choice may throw up too many artefacts of dubious quality to be helpful, but it is a possibility which may be helpful on occasions..

Obviously, it would be open to anyone to review an artefact and post their review as another artefact on AcademicArtefacts.org or elsewhere on the web. However, in practice this is likely to be the exception rather than the rule: Shashok and Matarese (2018) recount their difficulties in getting such a review published and claim that "although international organizations involved in science publishing support PPPR [post publication peer review], journals may not facilitate this type of scientific exchange, and institutions appear not to encourage it".

This means that there would be a need for the kind of organised reviewing and filtering process which is currently done by peer reviewed journals. Nosek and Bar-Anan (2012) use the phrase "review service" for such organisations, whereas Priem and Hemminger (2012) favour the word "assessment" which seems a more general formulation. Whatever term is used, it is likely that "quality judgments ... will be subtler than simple binary yes/no stamps" (Priem and Hemminger, 2012). These assessment organisations would take on the important role currently taken by journals and their editors of sourcing the reviews and deciding which papers to include in their collection; the important difference being that these assessment organisations are not restricted to papers submitted to one journal.

The Doi system (used by most journals and repositories) would have an important role to play in “allow[ing] things to be uniquely identified and accessed reliably” (doi.org accessed on 12 December 2023), as it is obviously important that readers are clear about the precise artefact that a review refers to.[6]

"Overlay journals" are a similar idea in that they ask authors to deposit documents on a repository (e.g. arxiv.org) and then perform the peer review function. These are a potentially valuable addition to the ecosystem. For example Quantum (https://quantum-journal.org/) "is a **non-profit** and **open access** peer-reviewed journal that provides high visibility for quality research on quantum science and related fields. It is an effort by researchers and for researchers to make science more open and publishing more transparent and efficient" (https://quantum-journal.org/about/ accessed on 6 April 2021). However, from my perspective the word "journal" is misleading in the sense that it is not a collection of articles: it is an overlay on articles posted elsewhere. And there are two further problems. First, it does not include articles published in other journals or repositories. Second, the assessment is restricted to reviews by peers.

Another possibility is represented by Qeios (https://www.qeios.com/) which boasts "AI-assisted invitation of the most suitable peers to review your first paper on Qeios", as well as a general encouragement to readers to review papers posted on the platform. The definition of “peer” here is obviously implicit in the AI which is, I assume, a black box, but this model does have the possibility of extending the review beyond peers chosen by an editor. I have published a paper on Qeios which I revised based on comments from 16 reviewers – see Appendix 4.

[6] Otherwise, there would always be the possibility that, for example, a paper presenting positive results of a drug trial might be substituted by another version of the paper in which the drug has been changed, perhaps to a cheaper one. This sort of deception could obviously have serious consequences.

Instead of papers being published in a peer reviewed journal which provides some guarantee of quality, my suggestion is that they should be published on a repository, and then their quality could be assessed by an assessment organisation. Then readers could search for artefacts of interest on a repository and then check how they are evaluated by the appropriate assessment service, or they could search for artefacts that have been evaluated by a particular assessment organisation. There are many possible models for these. Authors might apply to have artefacts assessed, or the assessment organisation might decide which artefacts to assess. Some of these assessment organisations might publish details of individual reviews; others might not. Conventionally peer reviewers are anonymous, but there are strong arguments in favour of publishing authors and details of reviews (Nosek and Bar-Anan, 2012: 236-8). In a more open system, if readers find open reviews helpful, there is likely to be competitive pressure to provide them. Some assessment organisations might stick with the current system of soliciting reviews from carefully selected experts, others might apply the principle of crowd sourcing and invite reviews from anyone.

This system would have the following advantages over the peer reviewed journal system:

1. Artefacts would be disseminated earlier because it would not be necessary to wait for a potentially lengthy review process (especially if a paper is submitted to a series of journals).
2. An assessment organisation could evaluate artefacts from any source including peer reviewed journals, whereas the peer review conducted by journals is restricted to articles published in the journals in question.
3. Artefacts could be evaluated by several assessment organisations. This would be helpful in enabling artefacts to be evaluated against different criteria and should provide a check on the verdicts provided by a given journal or assessment service. It would also make delays less likely if there are several assessment organisations involved. Many articles, of course, would not warrant detailed scrutiny by different experts using a range of criteria, but some articles would benefit. For example, one would want research on the safety and efficacy of covid vaccinations to be evaluated by a range of experts from a range of different perspectives. At present this doubtless happens informally, but the current peer review system does not encourage such a level of scrutiny.

As far as I know, no assessment organisations like this exist. However, it is an obvious idea, and would be easy to set up, so I there may well be such organisations which I do not know about.

The sections below flesh out a few possibilities that this "decoupled" system might facilitate.

# Possibilities that these principles might facilitate

## 1. Discipline or topic-based assessment organisations

Assessment organisations might be set up by learned societies interested in a particular discipline, or by people interested in a particular topic. For example, there might be one devoted to the Covid 19 pandemic which would review any work - wherever published - relevant to this topic. It would then publish a directory of credible research on this topic, with perhaps a note on any problems highlighted by reviewers. Like anything else, its credibility would depend on the credibility of the people or organisations behind it, but if such an organisation developed a good and deserved reputation, then it would provide a source for information on covid 19 and avoid the necessity to search separately for work on virology, epidemiology, behavioural science, statistics and so on.

## 2. Non-peer review: other perspectives and paradigms

Peer review is, of course, review by peers. This raises the question of why peers are the only people considered competent to review academic artefacts. Why not include non-peers in the assessment process? The basic case for non-peer review is outlined above.

Like non-fiction and non-vegetarian food in India[7], non-peer review might turn out to be a broader and more important category of review than it may initially seem. There are different categories of non-peers. Review from the perspective of lay readers is one possibility, although in many cases this would not be appropriate. Other possibilities are review from the perspective of neighbouring or competing disciplines, review from the perspective of relevant techniques such as statistical techniques, and so on. There are many possibilities, and it is not helpful to categorise them rigidly. Reviews from the perspective of another discipline should help to encourage the transfer of useful ideas across the boundaries of disciplines and sub-disciplines - something which the present system does not encourage. Some non-peer reviews might legitimately be regarded as less rigorous than peer reviews, but others might apply more rigorous criteria.

One potential benefit of non-peer review is that it may make it easier to challenge the dominant paradigm in a discipline. The present peer review system is almost certain to have a conservative influence: it is not likely to be receptive to new ideas. In the terminology of Thomas Kuhn's *The structure of scientific revolutions* (1970), peer reviewers are likely to favour "normal science" and to be sceptical about ideas which depart from received wisdom. Stanford (2019) argues convincingly that "historical transformations [which include peer review] of the scientific enterprise have generated steadily mounting obstacles to evolutionary, transformative, or unorthodox scientific theorizing, but also that we have substantial independent evidence that the institutional apparatus of contemporary scientific inquiry fosters an exceedingly and increasingly theoretically conservative form of that inquiry." Luukkonen (2012) puts forward a similar argument in relation to research funding. Horrobin (1990) gives many examples in the biomedical sciences where peer review has resulted in the rejection of important articles and the "suppression of innovation". This conformist tendency is likely to be reinforced by journal ranking systems which may make journals fearful of risking their status by accepting risky work.

Normal science, or following the conventional route, is, of course, often a good idea. Science progresses by following conventions, procedures and assumptions which are seen to work, but a system which does not allow established ideas to be challenged runs the risk of getting stuck in a blind alley (see Appendix 1 for a discussion of this problem in relation to the covid 19 pandemic). Fostering new ideas is difficult, almost by definition, but encouraging non-peer reviews seems more likely to achieve this than legitimising research just on the basis of reviews from two or three peers chosen by the editor of a journal. This might be a step towards Paul Feyerabend's "epistemological anarchy" (Preston, 2020), or the "disruptive science and technology" envisaged by Russell (2012).

[7] A reviewer of this paper queried this analogy. However, I think all three "non" categories are interestingly similar in that they are all based on an assumption about the main subcategory of something: most books are fiction, most food in India is vegetarian, and most useful reviews are by peers.

In my view, in years to come, the idea that the main way of assessing or legitimising academic research is via the views of two or three peers working in the same field will seem extremely odd. Surly a broader perspective can only be beneficial? Perhaps "peer review" should be replaced by "expert review" - which, of course, may encompasses other experts as well as peers.

## 3. A more critical approach to knowledge

The importance of being critical of ideas and theories is widely accepted. If a branch of knowledge is to progress, it is likely to be helpful to subject existing ideas to detailed critical scrutiny. For the philosopher Karl Popper (1972) testing and criticising scientific theories is crucial to the whole scientific enterprise[8]. There is a famous collection of articles from a meeting of philosophers of science (Lakatos and Musgrave, 1970) entitled "Criticism and the growth of knowledge": the title of the collection neatly summarises the spirit of many of the contributions.

Currently reviewing papers for journals is voluntary, unpaid and usually anonymous with comments and criticism not being published. In a system where the process of evaluating artefacts is separated from the decision to publish, it is possible that reviewing and evaluating research might have more status than it does currently, and the quality, variety, extent and visibility of the evaluations might be enhanced. There is also the possibility that some assessment services might pay reviewers for their services. Arguably, too much effort now goes into producing research, and not enough into evaluating and improving it.

## 4. Living artefacts incorporating corrections, retractions, and updates

According to Wikipedia "a living document, also known as an evergreen document or dynamic document, is a document that is continually edited and updated" (https://en.wikipedia.org/wiki/Living_document accessed on 23 April 2021). Journals publish articles with a publication date after which they are not changed. Updates, retractions, corrections and comments may appear on the web and in later issues of the journal, but readers may not see them. Repositories, however, often incorporate updates, different versions, corrections and so on. They can be, in effect, an evolving, or living, as opposed to a static, document.

This is obviously particularly welcome when important flaws are found in the research - readers of the paper can then be alerted to this. Appendix 1 describes one high profile example of faulty research published in the *Lancet* which was not fully retracted for 12 years. More recently, Covid 19 has obviously focused the *Lancet*'s editors' minds: a paper published in June 2020 was retracted within two weeks (Editors of the Lancet Group, 2020).

[8] For Popper scientific theories should be falsifiable in the sense that there are conceivable empirical observations which could prove them wrong. The idea of criticism is, however, broader than this: as well criticising a theory on the grounds that it's wrong, one might also criticise it on the grounds that it's too complicated, or its range is too restricted, or it's just not elegant. Also, theories whose conclusions are probabilistic make Popper's idea of falsifiability rather hazy.

Even without obvious flaws research may move on so that readers may benefit from seeing more recent work. One of my most cited papers is now 15 years old: I have now published another paper which takes the argument much further but it is the older paper which has the citations, so this is the one which readers find, not the recent update. Which is frustrating.

## 5. Different levels of assessment: true, possible, suggestion, etc?

Peer reviewed Journals publish papers that meet their peer review threshold and reject those that don't. There is no half-way house like "good suggestion but needs work", or "may be useful but we (the assessors) aren't sure". Assessments like these might assist the process of discussing possibilities and incubating ideas by publicising work without claiming it is the finished article. Perhaps the present article fits into this category?

## 6. Providing a home for ideas that don't fit existing journals

One of areas which interests me is the possibility of making knowledge simpler. (I will say a bit more about this in the next section.) Having written an article on this theme, I then looked for a suitable journal to submit it to but failed to find one. Simplifying knowledge did not seem to be a recognised research area. In the end I published it in an education journal (Wood, 2002), and focussed on the educational aspect of my thesis. However, the argument is actually much broader so publishing it in an education journal seriously restricted its scope and its audience. Several years later I wrote a more general article (Wood, 2016) and, as I had submitted several papers to arxiv.org which is "a free distribution service and an open-access archive for ... scholarly articles in the fields of physics, mathematics, computer science, quantitative biology, quantitative finance, statistics, electrical engineering and systems science, and economics," I tried submitting it to this repository. However, their list of fields does not include simplifying knowledge, so they duly rejected it. I subsequently submitted it to another repository, SSRN which did accept it, and several years later a revised version to Qeios (Wood, 2023; see Appendix 4). I also submitted version 1 of the present paper to SSNN and arxiv.org: arxiv.org accepted it but classified it as computer science, SSRN accepted it without giving it a subject classification. Classifying this paper as computer science seems unhelpful: the keywords below the abstract above give a better indication of its contents.

As time progresses fields of human knowledge will inevitably change: some fields will die out, new ones will emerge, and the character of many will evolve over time. Unlike the general repository suggested above, journals tied to specific fields are likely to exert a brake on such changes.

Peer review in a new discipline is also problematic because there are, by definition, no peers. Non-peer review may be the only option.

## 7. More emphasis on simplicity

Specialist disciplines evolve their own jargon and perspectives which workers in the field absorb as they learn their trade. The learning process may be cumulative in the sense that each step may require an understanding of previous steps. In extreme cases technical work at the leading edge of a discipline may be inaccessible to anyone outside the discipline. This creates two problems. First, reaching the leading edge may take a long time: it is possible to imagine that in years to come workers in some fields will need to spend so long reaching the leading edge that by the time they get there their cognitive powers are declining and they are thinking about retiring. This would mean that the discipline is stuck and cannot progress further. Second, inter-disciplinary collaborations and review of work by outsiders or non-peers of the kind I am arguing for here, is impossible. We will just have isolated islands of expertise, linked to nothing.

Most disciplines are a long way from either scenario at the moment, but they do illustrate the potential problems. The obvious way of reducing the seriousness of these problems is to try to make ideas - theories, models, jargon, etc - as simple as possible (but not, of course, simpler, as Einstein supposedly said), and as accessible to outsiders as possible. Editors of journals encourage authors to make the *presentation* of their ideas as simple as possible, but there is, in my experience, little to no emphasis on the simplicity of the ideas themselves. (This may, of course, not be true of all fields.) Sometimes, the opposite is the case: specialists may feel they need complex jargon and formalisms to enhance their status and keep outsiders out. In the long term, of course, this is self defeating; if simplicity is not taken seriously as a criterion for evaluating academic artefacts we are likely to end up drowning in a sea of half understood technicalities (see Wood, 2023). With luck, non-peer review, and less dependence on journals solely for specialist audiences, will encourage more emphasis on simplicity.

## Barriers to change

The change to the ecosystem I am suggesting could happen gradually, with the new systems co-existing with existing practices.

The present ecosystem is about half way to satisfying Principle 1 in that many, possibly most, academic artefacts are posted on open-access repositories such as ssrn.com and arxiv.org, and search engines such as Google Scholar can be used to find relevant work. However, the fact that many repositories are subject specific does complicate the process of both depositing and finding academic artefacts. It is possible that one global repository may become dominant in the same way that Google has become the dominant search engine, but the fact that some repositories do not allow submissions which are stored elsewhere is likely to inhibit this process.

In terms of Principle 2, there is nothing to stop professional bodies setting up their own reviewing service which would review relevant work from different sources. Instead of the Society for Bungee Jumping Science setting up their own journal, with all its associated costs, they could produce a compilation of reviews of artefacts considered relevant, and of a suitable quality, from a range of different sources.

However, resistance to change from some stakeholders would be inevitable. The next two sections discuss, briefly, financial issues, and the inertia of the present system, particularly concerning reputational factors.

## Financial costs and benefits

Old style journals are funded by selling the journals to readers and libraries, often at exorbitant prices. For obvious reasons, the publishers making money from these journals are likely to resist change. However, this regime is already starting to crumble with the advent of open access journals. At the moment most repositories have no charge for posting papers: arxiv.org for example is hosted by Cornell University and acknowledges "support from the Simons foundation and member institutions". Quantum, an overlay journal based on arxiv.org, suggests a voluntary contribution upon acceptance of a paper. Nosek and Bar-Anan (2012) review some of the possible funding models. It is worth bearing mind that the web does seem to foster large scale developments whose funding mechanism is not planned in advance: the absence of a clear business model is unlikely to discourage innovation in this area.

## Inertia, especially to do with reputation and research assessment

The main barrier to change in the market for academic knowledge is probably that researchers feel that their reputations, and salaries, depend on publishing in the top journals. This gives the top journals a monopoly on prestigious research. In time, papers published in repositories on the web (like arXiv.org) and reviewed elsewhere may achieve the necessary status to make the journals unnecessary, but at the moment in most disciplines the stranglehold of the top journals is still strong.

There is a tendency for crude and rigid grading systems to be imposed in markets where customers may have difficulties in choosing between competing products. So schools and hospitals are graded with some being labelled failures, consumer reports by organizations such as Which? (http://www.which.co.uk/) give scores to products and label some best buys. Similarly the pressure to grade academic departments in order to decide which are deserving of funding has led to pressures to grade research papers for which journal grading (four star, three star etc) is the common proxy used. This is unfortunate from many points of view - including the healthy growth of knowledge. Inevitably the grades will ignore some important criteria. In the sciences there is even evidence that retraction rates, due to fraud or errors in the research, are *higher* in the higher ranked journals (Brembs & Munafo, 2013). Academic research deserves a more flexible system for quality evaluation which the suggestions above should help to foster.

Despite this, the increasing pressure for open access to research papers may provide an escape route. Research published in the so-called top journals is increasingly likely to be made freely available - either in the journals themselves, or by depositing copies in institutional or subject-based repositories. And, of course, anyone is free to publish reviews of the papers in these journals. At the moment this is not done to any great extent because of the assumption that publication in a good journal is the only type of review necessary, and the barriers to publishing such reviews (see above) discussed by Shashok and Matarese (2018). However, in the future, particularly if links to freely available copies of the papers in question are available, such reviews could easily be published to provide more information than the mere fact of publication in a prestigious journal.

## Conclusions

I have discussed some suggestions to improve the academic knowledge market. At the moment, if I write an article on, say, the effectiveness of a new diet, I need to decide which peer reviewed journal to submit it to, make sure it conforms to the requirements of the chosen journal, send it off, wait for and react to the comments of the editor and reviewers, and then if it is rejected send it to another journal, and so on. This may take a long time, and the paper may get reviewed only from the perspective of the one journal - and it may be far from clear what sort of evaluation has been carried out.

Under the alternative system proposed here, I would post the article, or other academic artefact, on a public website - ideally a repository covering a broad range of subjects with a transparent search procedure. Then it might be reviewed from the statistical point of view by one assessment organisation, from the medical point of view by another, and from the social science point of view by a third. Another organization might produce a one paragraph summary in layman's terms of the status and credibility of my paper, assuming of course that it met their minimum standards - they might insist on, and draw on, reviews from the previous three bodies. Some quality stamps might depend on others: for example a medical reviewing service might insist on evaluation by a statistical reviewing service. Obviously many articles would not justify this level of scrutiny, but it could be invaluable in the case of, say, a medical article proposing an important but risky new treatment.

There are many potential benefits of such a system. Artefacts should be published quicker. Readers would be able to see how they have been evaluated on a range of criteria, including those of non-peers from outside the narrow confines of the academic tradition to which a paper belongs. The system should be more responsive to errors and updates and retractions would be more visible. Ideas that challenge dominant paradigms, or that don't fit into any existing subject specialism, should find it easier to gain a foothold. And the criterion of simplicity, which is arguably essential to progress, should have greater prominence.

## Acknowledgments

I am grateful to Ashraf Labib, Andreas Hoecht, Valerie Anderson, Alex Tymon and several anonymous reviewers for feedback on earlier versions of this article. Their comments were very different: this is both a strength of peer review, and one of its frustrations.

# Appendix 1: A few stories to illustrate problems with the current system

## The unreliability of peer review

If different reviewers from the pool of peers give different recommendations, then accepting the verdict of two or three of them makes the process into little more than a lottery. The two stories below illustrate this problem.

Peters and Ceci (1982) resubmitted 12 articles to the psychology journals which had published them 18 to 32 months previously, after changing the names of the authors and institutions and a few other minor details. Only three (25%) of the articles were recognised, and eight of the remaining nine were rejected by the same journals that had originally published them.

I first got into research on peer review when I asked a colleague who was organising a conference if I could have a look at the results of the peer reviews of the submitted papers (Wood et al, 2004). Each of the 58 papers was reviewed by two reviewers. When we divided the reviews into two categories - good and bad - 50% (29) of the papers received one good and one bad review. This percentage represented the proportion of papers where the reviewers did not agree; ideally the conference organizer would have preferred this percentage to be much lower. Overall 69 of the 116 reviews were good, and the rest bad. If these good or bad evaluations had been awarded to the papers at random the disagreement rate would have been 48%. In other words, the disagreement rate between individual reviewers of the same paper was slightly greater than it would have been if the process were entirely random. The reviewing process was, almost literally, a lottery. According to the conference organiser, one of the participants in the conference was decidedly unhappy because his paper had been rejected following two bad reviews. If the conference organizer had done this analysis at the time, he would surely have had less confidence in rejecting this paper.

The obvious response to this problem is to make the process more open (so that readers can see the basis and source of reviewers' judgments), and to publish, not reject, papers if there is any doubt - which is facilitated by the lack of space constraints on web-based systems.

## The consequences of erroneous decisions

Sometimes articles are published in prestigious journals and thus legitimised when they probably shouldn't have been. These range from accidental errors to systematic bias and fraud. One high profile example concerns an article published in the medical journal *The Lancet* in 1998 that

> "lent support to the subsequently discredited theory that colitis and autism spectrum disorders could be caused by the combined measles, mumps and rubella (MMR) vaccine. ... Investigations by Sunday Times journalist Brian Deer revealed that Wakefield [the author] had multiple undeclared conflicts of interest, had manipulated evidence, and had broken other ethical codes. The Lancet paper was partially retracted in 2004 and fully retracted in 2010 .... In 2011, Deer provided further information on Wakefield's improper research practices to the British medical journal, BMJ, which in a signed editorial described the original paper as fraudulent. The scientific consensus is that no evidence links the vaccine to the development of autism, and that the vaccine's benefits greatly outweigh its risks"
> (http://en.wikipedia.org/wiki/MMR_vaccine_controversy accessed on 3 September 2013).

The problem of articles like this is exacerbated by the fact that corrections may take a long time to be published (more than 10 years in this case), and that readers may not notice them when they are published because the original article, and the paper version of the journal in which was published, cannot be changed. With less serious errors, corrections may not be published and the original error, legitimized by publication in a prestigious journal, may persist and even be cited as an example of how things should be done. Nosek and Bar-Anan (2012: 237) give an example of two published articles in a high-profile journal which, they say, had critical errors in their analysis. One of these articles was cited by an editor of another journal as an example of how the analysis should be done. They comment that "it is very difficult to get such an error out of the present system."

There is obviously no foolproof way of avoiding error in research, but if reviews are available on the web, and articles are updated as problems are found, readers would be much more likely to be aware of criticisms. If these reviews were crowd sourced, as well as relying on two or three reviewers selected by an editor, there is a much greater likelihood that errors like this would be found.

The opposite problem, of course, occurs when an article is rejected when it perhaps should have been published. Horrobin (1990) outlines several such examples from the biomedical sciences.

## Problems arising in the covid 19 pandemic

Much of the research on the virus was published quickly on repositories such as medRxiv (e.g. Lavezzo, 2020) but with health warnings like "not certified by peer review". This is problematic because peer review cannot be relied on to ensure an article is correct. The peer review process itself has been made much quicker for Covid related research, although I read Lavezzo (2020) two weeks after it was posted and there was no note about any review despite the urgency of understanding the coronavirus.

More fundamentally, Greenhalgh (2020) asks "Will COVID-19 be evidence-based medicine's nemesis?" in an editorial in *PLOS Medicine*. She contrasted the "evidence-based medicine paradigm" with the "complex systems paradigm" and points out the rigidities of the former, with its emphasis on evaluating interventions by randomised controlled trials. The difficulty is that evidence which meets the high standards of the evidence-based medicine paradigm may be difficult or impossible to obtain. For example, she cited a report on the use of masks which was "criticised by epidemiologists for being 'non-systematic' and for recommending policy action in the absence of a quantitative estimate of effect size from robust randomized controlled trials." This ignores the obvious fact that even if the evidence on the use of masks is below the highest standards, it may still be a good idea to recommend their use because of the strong possibility that they may help prevent infections.

According to an article in *Nature* (Lewis, 2020) entitled "Mounting evidence suggests coronavirus is airborne — but health advice has not caught up", theories about the transmission of the coronavirus have clung to the idea that aerosols (small droplets in the air) are not an important route of transmission despite evidence to the contrary, which means that the health advice given by the supposed experts is deficient and may have cost many lives.

In both cases, a more flexible and multi-dimensional approach to evaluating research might help to reduce the influence of unhelpful dominant paradigms.

## Appendix 2: A very crude model of the speed of intellectual evolution

To give some idea of the potential impact of some of the problems of the reviewing process, consider a very crude probabilistic model dealing with the speed of intellectual evolution. Let's suppose that groundbreaking innovations in a particular field are produced, on average, every $n$ years, and that the probability of the movers and shakers (M&Ss) of the domain noticing them is $p$. This means, for example, that if groundbreaking research is produced every five years ($n = 5$) but only one in ten of these innovations are noticed ($p = 0.1$), then the mean time to the next innovation which is noticed is 50 years[9], 45 years longer than it would have been if the M&Ss noticed every groundbreaking innovation.

This model is obviously very crude but it does illustrate the key importance of the two quantities involved. How can we improve the values of two key figures in the model? The probability of a piece groundbreaking research being noticed (p) can be decomposed into the probability of publication ($p_p$) multiplied by the probability of it being noticed once it's been published ($p_{np}$). The first probability ($p_p$) should be as close to one as possible: the problem here is that of ensuring that the review system works as efficiently as possible and does not exclude useful innovations. The traditional way of improving the second probability ($p_{np}$) is to reduce the size of the pool of published papers; the difficulty here is that this might lead to the exclusion of useful innovations. With a larger pool of papers for researchers to sort through, the question is how can we improve the probability of groundbreaking innovations being noticed? Two ideas occur to me. First, the abstract is critical. Many researchers may only read the abstract so this needs to be very carefully formulated. Second, there may be a case for reading a random sample of the literature: this would mean that different researchers would read different samples which should increase the diversity of the ideas being carried forward, in contrast to the present scenario where everyone tends to cite the same works.

---

[9] Imagine a time span of 1000 years. The values of $n$ and $p$ (5 and 0.1) suggest about 200 innovations will be produced, of which 10%, or 20, will be noticed by the M&Ss - which is, on average one every 50 years.

The other variable is the frequency with which groundbreaking research is produced. I can envisage no appropriate, simple model to help understand this, but one reasonable assumption is that innovations are fostered by exposure to a variety of types of research which in turn is more likely with the open, system of dissemination I am proposing here. The censorship which is likely with the traditional system is likely to have the effect of inhibiting the development of a "long tail" (Anderson, 2006) of infrequently read, but potentially important, research articles.

## Appendix 3: Papers and other academic artefacts

The conventional unit of research output is the "paper" – a written report, on paper, of research carried out. Papers are typically about 2,000 - 15,000 words and include citations to other relevant work. There are different types of paper – review papers, short papers sometimes called letters, comments on other papers, etc – and the detailed conventions are different in different fields. However, they are all *written* documents, designed to be read by *peers*, with a clear *publication date* after which the documents are not changed, and a named list of *authors*. There may be good reasons to reconsider all these features.

With the advent of the web other formats besides the written word are obviously possible - videos, audio and pictures being the obvious examples, but there are other possibilities such as computer programs. Conventionally these are not used for presenting research, but this may change.

Conventional scholarly papers are intended for peers in the discipline. In practice the readership may be wider than this - researchers in other disciplines may be interested, as might members of the general public. This has implications for the prior knowledge assumed by the authors of the paper. Many conventional papers may give more background than is necessary for close peers, but insufficient for outsiders.

It is also worth pointing out the importance of named authors in academia. If you buy a car, or have a meal in a restaurant, you usually have no idea about the identities of the creative team behind the product. If you read some research, however, the author is an important and prominent piece of information. Academic pride and careers depend on authorship. This may lead to disputes, and to a failure to share data and ideas. If, in another parallel universe, authors of scholarly papers were anonymous figures in the background, would this make a difference to the progress of knowledge? It is impossible to know, but the success of Wikipedia suggests that something similar may be worth considering in academia (Rice, 2012). Obviously, there are important issues concerning intellectual property, copyright and patents, and the motivation of researchers here.

## Appendix 4: My experience posting a paper on Qeios.com and revising it based on comments from 16 reviewers

The first version posted on Qeios is at https://doi.org/10.32388/RHTT8D and the revised version is at https://doi.org/10.32388/RHTT8D.2. The claim on the website about the reviews helping to "improve [papers] in record time with the Open Peer Review from top experts in the field" was largely borne out by my experience, although the notion of "top experts" is not well defined in relation to my paper. I have summarized the improvements in an Appendix at the end of the second version of the paper (Wood, 2023).

Qeios uses an AI system to select suitable “peers” to ask to review papers: in the case of my paper, peers would have been difficult to find or even define as the paper does not fit into an established discipline (see the Section entitled Providing a home for ideas that don't fit existing journals above). The reviewers included a psychologist and physicists which is helpful given the topic of the paper. Qeios, I think, is consistent with Principle 1: Academic artefacts should be freely available, with minimal categorisation and evaluation, and a transparent search mechanism, and the reviewing service is a very useful extension, particularly in terms of the number of reviews (browsing the papers on Qeios the 16 reviewers of the first draft of my paper is a not untypical number). The rather looser definition of “peer” might also be an advantage (see the section entitled Peer review above).

However, it does not, and could not, embody Principle 2: Organisations for assessing academic artefacts are necessary, and there are advantages in separating these from organisations for dissemination because this would require separate organizations which are trusted as authoritative in a particular domain by readers.